\documentclass[lettersize,conference]{IEEEtran}
\usepackage{multirow}
\usepackage{glossaries}
\usepackage{adjustbox}
\usepackage{xcolor}
\usepackage{hyperref}
\usepackage{cleveref}
\usepackage{url}
\usepackage{booktabs}
\usepackage{pifont}
\usepackage{subfig}

\usepackage{balance}

\def\BibTeX{{\rm B\kern-.05em{\sc i\kern-.025em b}\kern-.08em
    T\kern-.1667em\lower.7ex\hbox{E}\kern-.125emX}}

\usepackage[switch]{lineno}

\definecolor{darkgreen}{rgb}{0,0.5,0}

\newcommand{\newcmready}[1]{\textcolor{black}{\textbf{}#1}}
\newcommand{\new}[1]{\textcolor{black}{\textbf{}#1}} 
\newcommand{\newo}[1]{\textcolor{black}{\textbf{}#1}} 

\newacronym{pp}{PP}{Partial Product}
\newacronym{lsb}{LSB}{Least Significant Bit}
\newacronym{dnn}{DNN}{Deep Neural Network}
\newacronym{ppu}{PPU}{Partial Product Unit}
\newacronym{fa}{FA}{Full Adder}
\newacronym{gemm}{GEMM}{General Matrix to Matrix Multiplication}
\newacronym{patqat}{PatQAT}{Pattern-based Quantization Aware Training}
\newacronym{sgd}{SGD}{Stochastic Gradient Descent}
\newacronym{tflite}{TFLite}{TensorFlow Lite}
\newacronym{ptq}{PTQ}{Post Training Quantization}
\newacronym{vmac}{VMAC}{Vector Matrix Multiplier Accelerator}
\newacronym{hdl}{HDL}{Hardware Description Language}
\newacronym{hls}{HLS}{High-Level Synthesis}
\newacronym{dma}{DMA}{Direct Memory Access}
\newacronym{ip}{IP}{Intellectual Property}
\newacronym{patproc}{PatProc}{Pattern Processing}
\newacronym{uram}{URAM}{Ultra RAM}
\newacronym{pdp}{PDP}{Power Delay Product}
\newacronym{lut}{LUT}{look-up table}
\newacronym{mred}{MRED}{Mean Relative Error Distance}
\newacronym{asic}{ASIC}{Application-Specific Integrated Circuit}
\newacronym{fpga}{FPGA}{Field-Programmable Gate Array}
\newacronym{qat}{QAT}{quantization-aware training}
\newacronym{pi1}{$\Pi1$}{OR-based Carry-Disregard Partial Product Unit}

\IEEEoverridecommandlockouts

\begin{document}
\newcommand{\ul}[1]{\underline{#1}}

\title{FAME: An \ul{F}PGA-Based Platform for \ul{A}pproximate \ul{M}ultipliers \ul{E}valuation with Pattern-Guided DNN Retraining}

\author{\IEEEauthorblockN{Rappy Saha\textsuperscript{1}, Nima Amirafshar\textsuperscript{2}, Jude Haris\textsuperscript{1}, Nima Taherinejad\textsuperscript{2}, and José Cano\textsuperscript{1}}
\IEEEauthorblockA{\textsuperscript{1}\textit{University of Glasgow, UK} \textsuperscript{2}\textit{Heidelberg University, Germany} \\
}

\thanks{© 2026 IEEE.  Personal use of this material is permitted.  Permission from IEEE must be obtained for all other uses, in any current or future media, including reprinting/republishing this material for advertising or promotional purposes, creating new collective works, for resale or redistribution to servers or lists, or reuse of any copyrighted component of this work in other works.Accepted at SBAC-PAD 2026.}
}

\maketitle
\glsresetall

\begin{abstract}
\bstctlcite{IEEEexample:BSTcontrol}
Approximate multipliers can reduce hardware area and energy consumption in \gls{dnn} inference; however, they introduce computational errors. Assessing the accuracy of numerous approximate multiplier designs across diverse DNN models and large-scale datasets remains challenging due to prohibitive evaluation times. This overhead primarily stems from the slow emulation of approximate multiplier behavior using \glspl{lut} on CPU and GPU platforms. Moreover, the resulting accuracy degradation must be carefully quantified and, if necessary, mitigated (e.g., through retraining), further increasing the overall evaluation cost.

To address these challenges, we propose \texttt{FAME}, an FPGA-based platform for evaluating approximate multipliers. The platform exploits the reconfigurable logic of Field-Programmable Gate Arrays (FPGAs) to implement approximate multipliers directly in hardware, eliminating the need for \gls{lut}-based emulation on CPU/GPU platforms and thereby enabling efficient DNN inference while significantly reducing evaluation time on large datasets. Furthermore, we introduce a pattern-guided \gls{dnn} retraining technique to mitigate accuracy degradation induced by approximate multipliers. Specifically, retraining is guided by multiplier-specific patterns to effectively recover potential accuracy losses. 
We evaluate \texttt{FAME} using two DNN models, ResNet-18 and MobileNetV2, on the ImageNet dataset across \new{27} approximate multipliers. During inference, our approach achieves up to a \new{3.47$\times$} speedup in approximate multiplier evaluation compared to prior \gls{lut}-based emulation methods. Furthermore, the proposed retraining technique improves accuracy by up to \newo{65.5\%} over existing retraining approaches for the evaluated multipliers. The code is publicly available at: \url{https://github.com/gicLAB/FAME}
\end{abstract}

\begin{IEEEkeywords}
Approximate Computing, DNNs, FPGAs, Hardware-Software Co-Design, Edge AI.
\end{IEEEkeywords}

\begin{table*}[t]
    \centering
    \caption{Comparison of retraining and inference evaluation methods for approximate multipliers.}
    \label{tab:comparison}
    \resizebox{\textwidth}{!}{
    \begin{tabular}{c|c|c|c|c|c|c|c}
    \hline
    \multirow{2}{*}{\textbf{Work}} & \multicolumn{4}{c|}{\textbf{Retraining}} & \multicolumn{3}{c}{\textbf{Inference}} \\
    \cline{2-8}
    & \textbf{Quantization} & \textbf{Multipliers} & \textbf{Framework} & \textbf{Hardware} & \textbf{Multipliers} & \textbf{Framework} & \textbf{Hardware} \\
    \hline
    AdaPT~\cite{danopoulos_adapt_2023} & AMi8 & LUT AMi8 & PyT & CPU & LUT AMi8 & PyT & CPU \\
    TFApproxIL~\cite{pinos_acceleration_2023} & AMi8 & LUT AMi8 & TF & GPU & LUT AMi8 & TF & GPU \\
    TER~\cite{yu_toward_2024} & AMi8 (Exact region) & EXi8 & PyT/TF & GPU & EXi8 & PyT/TF & GPU \\
    Ours & AMi8 (Exact+Approx region) & EXf32 & PyT/TF & GPU & AMi8 & TFLite & FPGA  \\
    \hline
    \end{tabular}
    }
\vspace{0pt}
\begin{flushleft}
\footnotesize{EXf32:Exact 32-bit floating-point multiplier; PyT: PyTorch; TF: TensorFlow.}
\end{flushleft}
\end{table*}

\section{Introduction}

Quantization is a widely adopted technique for reducing the size of deep neural networks (\glspl{dnn}), thereby improving latency and energy efficiency, particularly for deployment on resource-constrained edge devices~\cite{DLAS_TACO2025}. In practice, 8-bit integer (INT8) quantization has become an industry standard~\cite{tflite}, offering approximately a $4\times$ reduction in model size compared to 32-bit floating-point representations. Consequently, the dominant computation in DNN inference shifts from 32-bit floating-point multiplications to more efficient 8-bit integer operations, yielding substantial gains in performance and energy efficiency. 
Building on this trend, recent work has explored approximate INT8 multipliers (AMi8) to further enhance hardware efficiency~\cite{Nima-PRIM,Salar-ACE,mrazek2017evoapprox8b}. Compared to exact INT8 multipliers (EXi8), AMi8 designs can significantly reduce hardware area, with the degree of savings depending on the approximation strategy employed. Given that multiplication is the primary computational bottleneck in DNNs, integrating AMi8 into custom hardware accelerators presents a promising opportunity to further improve performance and energy efficiency. 
\newcmready{However, the approximation errors introduced by AMi8 designs degrade the accuracy of INT8-quantized DNNs~\cite{danopoulos_adapt_2023}.}
Evaluating this degradation requires executing DNN models on AMi8-based hardware. Since such hardware is not readily accessible, prior work has relied on \gls{lut}-based emulation of AMi8 on CPU/GPU platforms~\cite{danopoulos_adapt_2023, vaverka_tfapprox_2020}. Using frameworks such as PyTorch~\cite{paszke2017pytorch} and TensorFlow~\cite{abadi2016}, these approaches simulate AMi8 behavior to estimate accuracy loss. However, this emulation incurs substantial computational overhead, even with optimizations~\cite{pinos_acceleration_2023}, and becomes a major bottleneck when evaluating multiple AMi8 designs across different DNNs and large-scale datasets (e.g., ImageNet~\cite{krizhevsky2012imagenet}).

To address this limitation, we propose \texttt{FAME}, an FPGA-based platform for approximate multiplier evaluation. \texttt{FAME} eliminates LUT-based emulation overhead by implementing approximate multipliers directly on the FPGA fabric. Specifically, AMi8 units are integrated into an FPGA-based GEMM accelerator and interfaced with TensorFlow Lite (TFLite), enabling efficient evaluation of INT8-quantized DNNs. This design significantly reduces evaluation time compared to conventional emulation-based approaches. 

In addition, to mitigate the accuracy loss introduced by AMi8, prior work has explored \newcmready{INT8} quantization-aware training (QAT) techniques that incorporate AMi8 behavior into the training loop~\cite{danopoulos_adapt_2023,pinos_acceleration_2023}. However, these methods still depend on LUT-based emulation, making retraining computationally expensive. To overcome this, TER~\cite{yu_toward_2024} proposed a retraining approach based on identifying exact regions of approximate multipliers. Specifically, it constrains \gls{dnn} weights (i.e., one operand of the multiplier) to \newcmready{INT8} values for which the AMi8 produces exact results \newcmready{(i.e., exact regions)}, ensuring error-free computation regardless of the \newcmready{INT8 values of the} other operand. This enables retraining using EXi8, thereby avoiding LUT-based emulation. While effective for certain AMi8 designs, this approach does not generalize to all \newcmready{AMi8} types.

To address this limitation, we propose Pattern-guided Quantization-aware Training (\texttt{PatQAT}). Unlike TER~\cite{yu_toward_2024}, our method constrains \newcmready{\gls{dnn}} weights to a set of \newcmready{INT8} values (referred to as patterns) that may produce both exact and approximate results for a given AMi8. Retraining is still performed using exact multipliers to avoid emulation overhead; however, the selected patterns include values that induce approximation errors. The resulting accuracy impact is then evaluated using \texttt{FAME} \newcmready{for that AMi8}. Experimental results show that \texttt{PatQAT} effectively improves DNN accuracy across various AMi8 designs, despite partially constraining weights to regions that introduce approximation.

Finally, we note that \texttt{FAME} is orthogonal to the retraining methodology. Existing approaches, such as LUT-based emulation~\cite{danopoulos_adapt_2023,pinos_acceleration_2023} or TER~\cite{yu_toward_2024}, can still be employed during retraining, while \texttt{FAME} can be independently used for fast and efficient inference evaluation.
To the best of our knowledge, \texttt{FAME} is the first open-source FPGA-based evaluation platform designed specifically for approximate multipliers, providing a significant advantage when evaluating \newcmready{large} \glspl{dnn} on large-scale datasets \newcmready{across many AMi8 designs}.

The main contributions of this work are as follows:

\begin{itemize}
    \item We propose \texttt{FAME}, an FPGA-based platform that accelerates the \newcmready{accuracy} evaluation of \glspl{dnn} using various approximate multipliers.

    \item We introduce \texttt{PatQAT}, a pattern-guided \gls{dnn} retraining method to recover accuracy loss caused by approximate multipliers.

    \item We demonstrate the effectiveness of \texttt{FAME} by evaluating ResNet-18 and MobileNetV2 on the ImageNet dataset with \new{27} approximate multipliers, achieving up to \new{$3.47\times$} speedup in evaluation and up to \newo{65.5\%} accuracy recovery using \texttt{PatQAT} compared to prior work.
\end{itemize}

\section{Background and Related Work}

\subsection{Approximate Multipliers}

Conventional exact multipliers suffer from long critical path delays and high power consumption, which, due to their frequent use in many applications, lead to increased overall energy consumption.
In general, an exact multiplier has three main stages: (1) \gls{pp} generation, (2) \gls{pp} accumulation, and (3) final addition. Among these stages, \gls{pp} accumulation has the highest hardware complexity.
In recent years, many approximate multipliers have been proposed using different approximation techniques.
The Booth architecture~\cite{Liu-booth, Booth2, survey1} is a common approach that reduces the number of \gls{pp} rows by using PP encoding. 
Another well-known structure is the logarithmic multiplier~\cite{Yin-log,survey1}, where Mitchell’s algorithm converts multiplication into addition, which reduces the circuit complexity.
Another promising approximation method is operand truncation~\cite{drum-iccad,survey1}, where the \glspl{lsb} of the operands are ignored, and only the remaining higher-order bits are used for multiplication using a smaller core multiplier. 

Tree-based multipliers~\cite{jiang2020approximate} improve the speed of \gls{pp} accumulation by using compressors as the main building blocks.
Although these multipliers have a shorter critical path delay, they often suffer from high power consumption and large circuit area~\cite{Nima-TCAS23}.
To address this issue, various approximate compressors have been developed and widely studied in the literature~\cite{invcam-electronics26,survey1}. 

In contrast, array multipliers have a uniform and modular structure, which can lead to lower area and energy consumption compared to compressor-based multipliers. 
However, the main bottleneck in array multipliers is carry propagation, which is the main reason for their high critical path delay. 
One effective approximation approach is carry disregard~\cite{Nima-PRIM, Nima-TCAS23, Nima-DSD22}.
By ignoring carries in some \gls{pp} columns of an array multiplier, these approximate columns can operate in parallel, which improves both delay and power consumption.

In this paper, to describe our proposed \texttt{PatQAT} and \texttt{FAME}, we focus on approximate INT8 multipliers (AMi8) and select 20 designs from the SPRIM8 family, a class of carry-disregard multipliers which is the signed version of PRIM8 multipliers~\cite{Nima-PRIM}. 
\new{We also evaluate 7 AMi8 from the EVOApprox family~\cite{mrazek2017evoapprox8b}, which are based on an evolutionary algorithm and have been widely studied in the previous literature~\cite{pinos_acceleration_2023,yu_toward_2024}.
Including both SPRIM8 and EVOApprox families in our evaluation allows us to demonstrate the effectiveness of our proposed methods across a diverse set of AMi8 designs.}


\begin{figure*}[t]
    \centering    
    \subfloat[]{
        \includegraphics[width=0.45\textwidth]{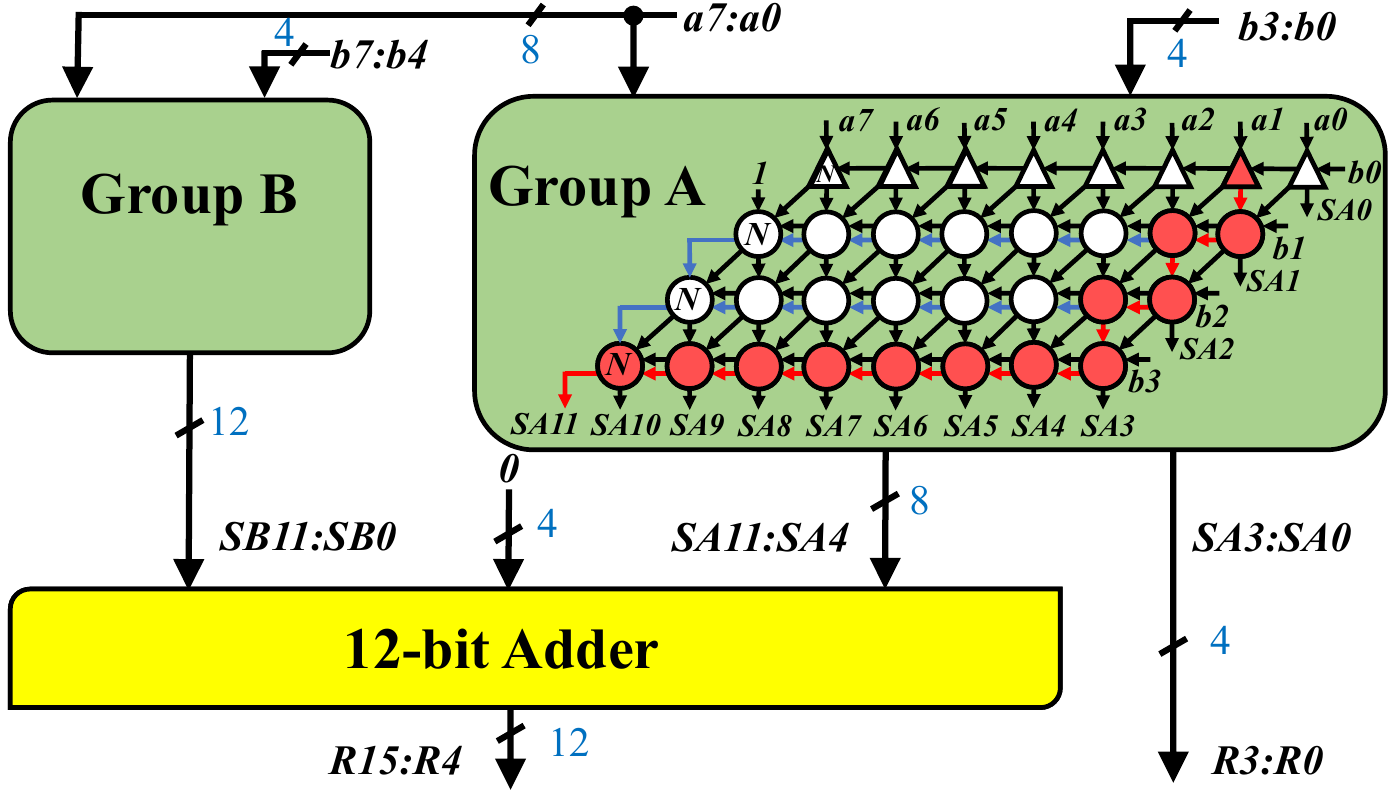}
    }
    \hfill
    \subfloat[$\Pi1\_1$]{
        \includegraphics[width=0.18\textwidth]{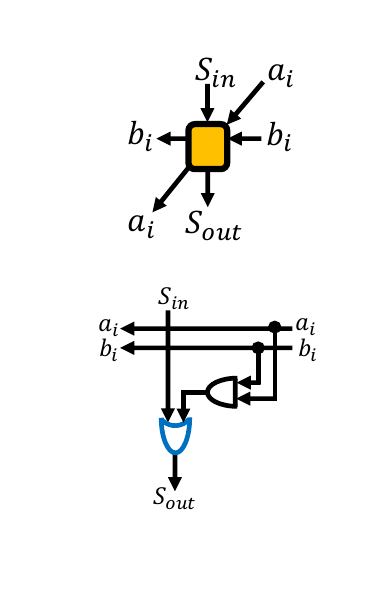}
    }
    \hfill
    \subfloat[$\Pi1\_2$]{
        \includegraphics[width=0.16\textwidth]{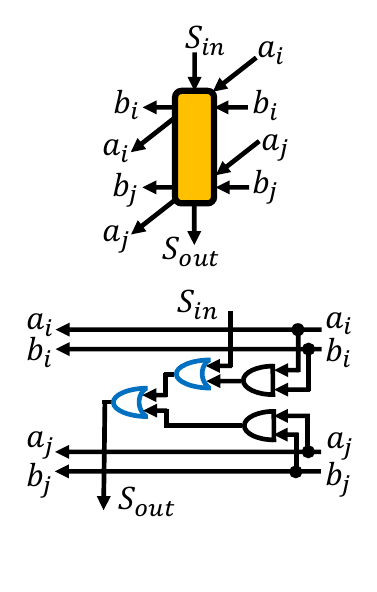}
    }
    \hfill
    \subfloat[$\Pi1\_3$]{
        \includegraphics[width=0.15\textwidth]{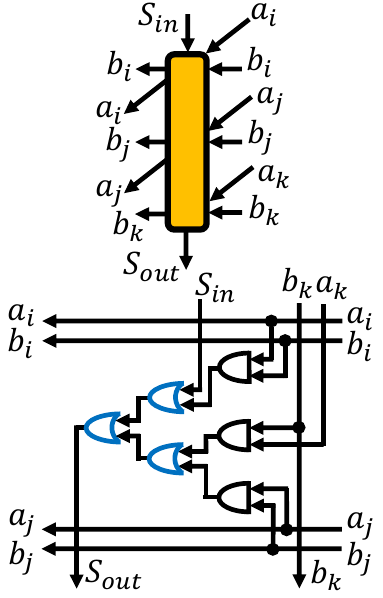}
    }
    \caption{(a) Exact 8-bit signed multiplier using two $8\times4$ groups; (b)-(d) Circuits of approximate partial product units ($\Pi1$s).}
    \label{fig:combined_em_ppu}
\end{figure*}

\subsection{DNN Retraining and Inference}

Previous studies have investigated \gls{dnn} retraining techniques to recover the accuracy loss introduced by approximate multipliers during inference~\cite{ansari_improving_2020,kim_effects_2022,gong_approxtrain_2023,pinos_acceleration_2023,danopoulos_adapt_2023,trommer_fast_2024,yu_toward_2024,meng_gradient_2025}. 
One common approach involves explicitly incorporating approximate multipliers into the training loop.
Several works provide native support for approximate multiplier-aware retraining and inference within popular deep learning frameworks, such as TensorFlow~\cite{de_la_parra_proxsim_2020,vaverka_tfapprox_2020,pinos_acceleration_2023,gong_approxtrain_2023} and PyTorch~\cite{danopoulos_adapt_2023,trommer_fast_2024} on CPU/GPU platforms. 
Other studies focus on developing retraining strategies that improve robustness and accuracy in the presence of approximate arithmetic~\cite{hammad_deep_2019,riaz_caxcnn_2020,ansari_improving_2020,kim_effects_2022,meng_gradient_2025}.
In addition to retraining-based approaches, some works evaluate the effectiveness of approximate multipliers in \glspl{dnn} without retraining~\cite{mrazek_alwann_2019,Salar-ACE}. 
ALWANN~\cite{mrazek_alwann_2019} identifies the most suitable approximate multiplier for each layer of a DNN, while ACE-CNN~\cite{Salar-ACE} applies post-training quantization (\gls{ptq}) on \glspl{dnn} using a customized TensorFlow graph. 
Across all these approaches, approximate multipliers are typically emulated using \gls{lut} on CPU/GPU platforms, while TER~\cite{yu_toward_2024} proposes a retraining and inference evaluation method using exact multipliers even when the target hardware uses approximate multipliers. 

In contrast to the LUT-based emulation and TER method, we propose: i) \texttt{FAME}, an FPGA-based platform for approximate multiplier evaluation during inference; and ii) \texttt{PatQAT}, a pattern-guided retraining method to recover the accuracy loss caused by approximate multipliers.
The key advantage of \texttt{FAME} is that it eliminates the emulation overhead, while \texttt{PatQAT} allows retraining using exact multipliers. 
Table~\ref{tab:comparison} shows the comparison of our proposed methods with prior works.
Our proposed methods enable researchers to efficiently explore the design space of approximate multipliers and their impact on \gls{dnn} accuracy, while also providing an effective solution for mitigating accuracy loss through retraining.

\section{Approximate Hardware-DNN Retraining Co-Design}

\newcmready{In this section, we first describe the design of the two classes of AMi8: SPRIM8 and EVOApprox (Section~\ref{sec:ami8}).
We then analyze the patterns (including approximate and exact regions) of the AMi8 designs in these two classes (Section~\ref{sec:pattern_analysis}).
Finally, we describe our proposed pattern-guided retraining method, \texttt{PatQAT}, which uses the analyzed patterns to retrain a \gls{dnn} for an AMi8 (Section~\ref{sec:patqat}).}


\subsection{Approximate Multipliers Design Selection}
\label{sec:ami8}



We now describe the circuits of carry-disregard multipliers, namely SPRIM8, which are selected \new{as one of the classes of AMi8 in this work.}
SPRIM8 designs are the signed version of PRIM8 multipliers, originally introduced in~\cite{Nima-PRIM}.
\new{We describe the structure of SPRIM8 to provide insights into pattern generation for carry-disregard class of approximate multipliers.
As a widely studied class of approximate multipliers, EVOApprox multipliers are also included in our evaluation, and we refer interested readers to~\cite{mrazek2017evoapprox8b} for their architectural details.}

\Cref{fig:combined_em_ppu}(a) shows the structure of an exact 8-bit signed multiplier, which serves as the baseline for SPRIM8 designs.
This structure consists of two independent groups of exact \glspl{ppu}, namely groups A and B, each accumulating four rows of \glspl{pp}.
In addition, a 12-bit Ripple Carry Adder (RCA) produces the final multiplication result. 
Group A and Group B follow an array structure in which each \gls{ppu} includes an AND gate to generate a \gls{pp} and a \gls{fa} to sum the \glspl{pp}. 
In general, carry propagation between \glspl{ppu} is the main bottleneck in this structure and leads to a long critical path delay (highlighted in red in \Cref{fig:combined_em_ppu}(a)). 
To address this challenge, all carries in SPRIM8 multipliers are disregarded from the first column up to a selected column in both groups, allowing the approximate columns to operate independently and in parallel.
In these carry-disregard columns, the conventional \glspl{ppu} are replaced with \glspl{pi1}.
\Cref{fig:combined_em_ppu}(b)-(d) show the circuits of the three types of \gls{pi1} units, which have neither carry input nor carry output; therefore, \glspl{fa} are not required.
In all \gls{pi1} units, \glspl{fa} are replaced by OR gates to generate the sum bits, which improves hardware metrics, including area, power, and delay.
Notably, using OR gates instead of XOR gates for sum generation not only improves hardware complexity but also reduces the error caused by carry disregard without requiring an additional compensation unit, as described in~\cite{Nima-PRIM}.
\gls{pi1}\_1 and \gls{pi1}\_2 are used in carry-disregard columns with two and three \glspl{pp}, respectively, while \gls{pi1}\_3 is an approximate 4:1 compressor used in columns with four \glspl{pp}.
For example, if carries are disregarded from column 1 to 4 in Group A (\Cref{fig:combined_em_ppu}(a)), then the exact \glspl{ppu} in columns 2, 3, and 4 are replaced with \gls{pi1}\_1, \gls{pi1}\_2, and \gls{pi1}\_3, respectively.

SPRIM8 has two subclasses: SPRIM8(R12), which uses an exact 12-bit adder, and SPRIM8(R10), which uses an approximate 12-bit adder. 
In SPRIM8(R10), the OR-based carry-disregard approach is applied to the first two bits, while the remaining bits are summed using an exact 10-bit adder. 
We select 10 multipliers from each subclass (20 multipliers in total).
As for the naming convention, we name the SPRIM8 class approximate multipliers as SP$xy$, where parameters $x$ and $y$ represent the number of carry-disregard columns in groups A and B, respectively.
\new{In case of the EVOApprox class, we use the original names of signed 8-bit multipliers(MUL8S) as provided in~\cite{mrazek2017evoapprox8b}}. 
The last column of Table~\ref{tab:pattern_analysis} shows the selected AMi8 from each subclass.

\begin{table}[t]
\Large
\centering
\caption{\new{Pattern Analysis for SPRIM8 and EVO~\cite{mrazek2017evoapprox8b} Approximate Multipliers.}}
\label{tab:pattern_analysis}
\noindent\makebox[\columnwidth][l]{%
\resizebox{\columnwidth}{!}{%
\begin{tabular}{ccccccc}
\hline
\multirow{2}{*}{\textbf{Pattern}} & \multicolumn{2}{c}{\multirow{2}{*}{\textbf{Exact Region}}} & \textbf{Approx} & \multicolumn{2}{c}{\multirow{2}{*}{\textbf{Pattern Region}}}  & \multirow{2}{*}{\textbf{AMi8}} \\
 &  & & \textbf{Region} &  & & \\
\hline
\multicolumn{7}{c}{\textbf{SPRIM8(R12)}} \\
\hline
$\rho$1 & [-128,120] & 80 & 127 & [-127,127] & 80 & SP41, SP42, SP43 \\
$\rho$2 & [-128,113] & 56 & 127 & [-127,127] & 56 & SP51, SP52, SP53 \\
$\rho$3 & [-128,113] & 32 & 127 & [-127,127] & 32 & SP44, SP54       \\
$\rho$4 & [-128,97]  & 16 & 127 & [-127,127] & 16 & SP45, SP55       \\
\hline
\multicolumn{7}{c}{\textbf{SPRIM8(R10)}} \\
\hline
$\rho$5 & [-128,120] & 56 & 127 & [-127,127] & 56 & SP41, SP42, SP43 \\
$\rho$6 & [-128,112] & 40 & 127 & [-127,127] & 40 & SP51, SP52, SP53 \\
$\rho$7 & [-128,112] & 20 & 127 & [-127,127] & 20 & SP44, SP54       \\
$\rho$8 & [-128,96]  & 12 & 127 & [-127,127] & 12 & SP45, SP55       \\
\hline
\multicolumn{7}{c}{\textbf{EVO(MUL8S)}} \\
\hline
$\rho$9  & [-128,126] & 128 & -127,127 & [-127,127] & 129 & 1KVM \\
$\rho$10 & [-128,124] & 64  & -127,127 & [-127,127] & 65  & 1KV8 \\
$\rho$11 & [-128,120] & 32  & -127,127 & [-127,127] & 33  & 1KV9, 1KVP \\
$\rho$12 & [-128,112] & 16  & -127,127 & [-127,127] & 17  & 1KVQ, 1KVA \\
$\rho$13 & [-128,96]  & 8   & -127,127 & [-127,127] & 9   & 1KX5 \\

\hline
\end{tabular}%
}%
}
\vspace{0.4em}

\noindent\makebox[\columnwidth][l]{%
\resizebox{\columnwidth}{!}{%
\begin{tabular}{ccccccc}
\hline
\multicolumn{1}{c|}{\multirow{2}{*}{\shortstack{Formation of \\$\rho$8}}} & 
\multicolumn{6}{c}{Exact Region:  [-128, -127, -96, -64, -63, -32, 0, 1, 32, 64, 65, 96]} \\
\multicolumn{1}{c|}{} & 
\multicolumn{6}{c}{Pattern:  [-127, -96, -64, -63, -32, 0, 1, 32, 64, 65, 96, 127]} \\
\hline
\multicolumn{1}{c|}{\multirow{2}{*}{\shortstack{Formation of \\$\rho$13}}} & 
\multicolumn{6}{c}{Exact Region:  [-128, -96, -64, -32, 0, 32, 64, 96]} \\
\multicolumn{1}{c|}{} & 
\multicolumn{6}{c}{Pattern:  [-127, -96, -64, -32, 0, 32, 64, 96, 127]} \\
\hline
\multicolumn{7}{c}{$\rho$1$\subset$$\rho$2$\subset$$\rho$3$\subset$$\rho$4 \qquad $\rho$5$\subset$$\rho$6$\subset$$\rho$7$\subset$$\rho$8 \qquad $\rho$9$\subset$$\rho$10$\subset$$\rho$11$\subset$$\rho$12$\subset$$\rho$13} \\
\hline

\end{tabular}%
}%
}
\vspace{-6pt}
\begin{flushleft}
\scriptsize{Exact Region(Range, Size); Pattern Region(Range, Size).}
\end{flushleft}
\end{table}


\subsection{Pattern Analysis}
\label{sec:pattern_analysis}

For an AMi8 multiplier, which has two INT8 inputs, the range for each input is from -128 to 127, that is, \emph{256} possible values for each input.
To find the exact region of an AMi8, we focus on one INT8 input and analyze the output for all possible values regardless of the values applied to the other INT8 input.
For the set of INT8 values in the first input that produce the exact output regardless of the values in the second input, we define this set as the exact region of an AMi8 (see Table~\ref{tab:pattern_analysis}).
For the rest of the INT8 values in the first input that produce approximate output, we define them as the approximate region of an AMi8 (Table~\ref{tab:pattern_analysis}).
In Table~\ref{tab:pattern_analysis}, the exact region is represented as a range, and the AMi8s that produce the same exact region are grouped together.
For example, in the SPRIM8 (R12) class, the SP41, SP42, and SP43 approximate multipliers share the same exact region, defined over the range $[-128, 120]$.
This exact region consists of \emph{80} INT8 values within the range, for which these multipliers produce error-free (exact) results.
\new{To form the pattern of an AMi8 multiplier, we focus on including the -127 and 127 INT8 values in the pattern, which are the maximum negative and positive for symmetric INT8 quantization~\cite{jacob_2018_CVPR}, respectively.
We also remove the -128 INT8 value from the pattern, which is not required for symmetric INT8 quantization.
In Table~\ref{tab:pattern_analysis}, we present examples illustrating the formation of patterns $\rho$8 and $\rho$13, which are the smallest patterns in the SPRIM8 and EVOApprox classes, respectively.
For pattern $\rho$8, we take all INT8 values in the exact region except -128, and take 127; whereas for pattern $\rho$13, we take all INT8 values in the exact region except -128, and take both -127 and 127 from the approximate region as the pattern range.
}
To find the patterns for each AMi8 multiplier, we create a Python behavioral model and analyze the output for all possible combinations of INT8 inputs.

\new{
The SPRIM8(R12) and SPRIM8(R10) classes of approximate multipliers produce eight unique patterns ($\rho$1 to $\rho$8), and the EvoApprox class produces another five unique patterns ($\rho$9 to $\rho$13).
}
The variation in patterns between the two subclasses \new{of the SPRIM8 class} is due to the use of an exact 12-bit adder in the final addition stage of SPRIM8(R12) and an approximate 12-bit adder in the final addition stage of SPRIM8(R10).
Besides, as the approximation increases with the number of carry-disregard columns (SP$xy$), the size of the exact region as well as the pattern decreases.
From our observation, we find that patterns are hierarchical in nature within each class, i.e., $\rho$1$\subset$$\rho$2$\subset$$\rho$3$\subset$$\rho$4 for SPRIM8(R12), and $\rho$5$\subset$$\rho$6$\subset$$\rho$7$\subset$$\rho$8 for SPRIM8(R10). 
This hierarchical structure also stems from the number of carry-disregard columns.
Due to this hierarchical nature of patterns, the smallest pattern exists in all approximate multipliers of a class.
For example, pattern $\rho$4 exists in all approximate multipliers of the SPRIM8(R12) class and
pattern $\rho$8 exists in all SPRIM8(R10) approximate multipliers.
\new{
Although the EVOApprox class AMi8 multipliers are based on an evolutionary algorithm, we also observe a similar hierarchical structure in their patterns, i.e., $\rho$9$\subset$$\rho$10$\subset$$\rho$11$\subset$$\rho$12$\subset$$\rho$13.
}

The key difference between TER~\cite{yu_toward_2024} and \texttt{PatQAT} is that TER defines the pattern as the exact region of an AMi8 multiplier, while we define the pattern as a combination of both exact and approximate regions of an AMi8.
If we consider the exact region as the pattern, as in TER, for our selected AMi8, the accuracy recovery is limited (see Section~\ref{sec:compare_retraining}).
Therefore, our proposed pattern generalizes the retraining method to wider classes of AMi8.


\begin{figure}[t]
    \centering
        \includegraphics[width=0.95\columnwidth]{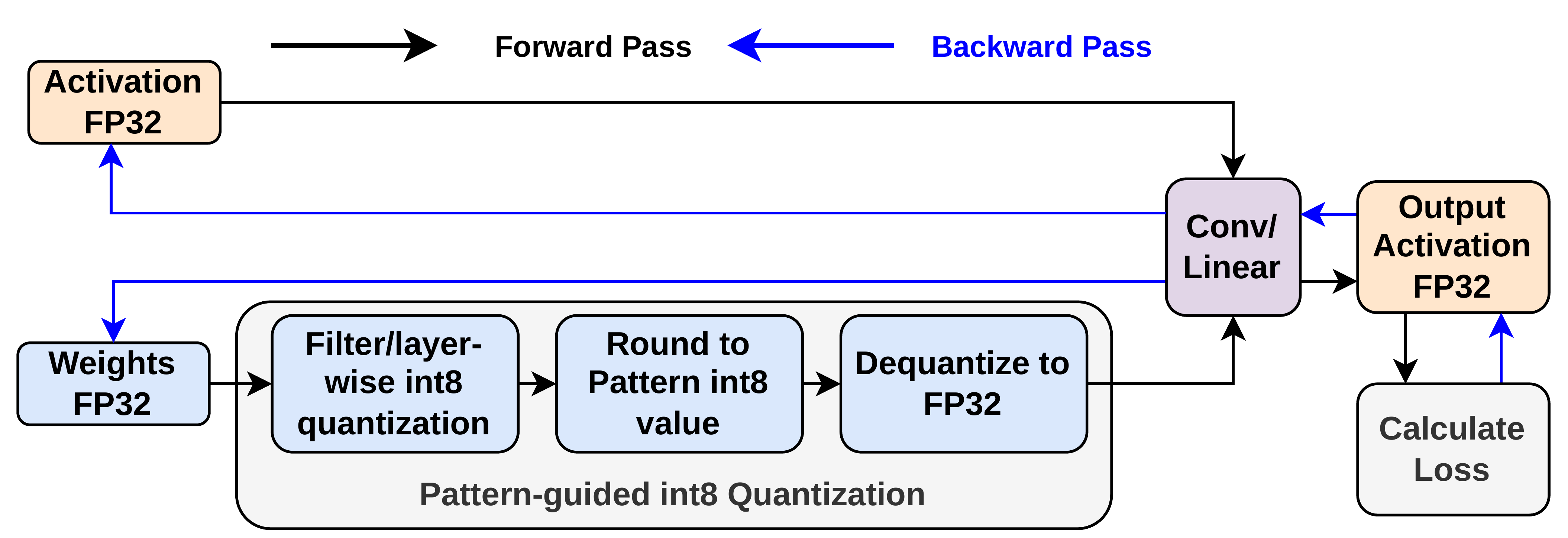}
    \caption{Forward and Backward Pass of \texttt{PatQAT}.}
    \label{fig:patqat_framework}
    \vspace{-1em}
\end{figure}

\begin{figure*}[t]
    \centering
        \includegraphics[width=\textwidth]{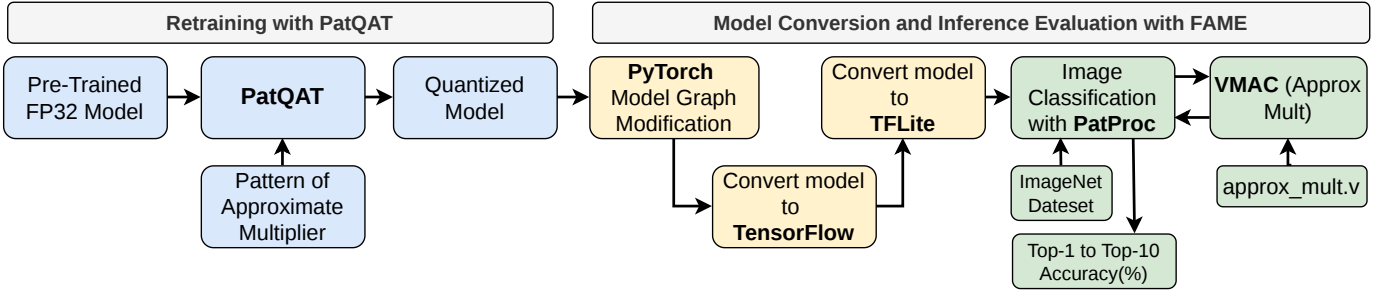}
    \caption{\newcmready{DNN inference evaluation with \texttt{FAME} (green), retraining with \texttt{PatQAT} (blue), and model conversion (yellow).}}
    \label{fig:proposed_framework}
\end{figure*}

\subsection{Pattern-guided Retraining (\texttt{PatQAT})}
\label{sec:patqat}

The purpose of retraining a \gls{dnn} with the analyzed patterns (Table~\ref{tab:pattern_analysis}) is to adapt the model weights within the defined patterns so that during inference, when approximate multipliers are used, the accuracy degradation is minimized. 
Additionally, retraining fine-tunes the model for the patterned weight to further improve the accuracy.
To do the retraining, the PyTorch~\cite{paszke2017pytorch} framework is used.
The pre-trained floating-point 32-bit (FP32) models from Torchvision~\cite{pytorchvision} are used as the starting point for retraining.
The forward and backward passes of the retraining process are shown in Figure~\ref{fig:patqat_framework}.
During the retraining process, we kept activations in FP32 format. 
Weights are quantized to patterned INT8 values and dequantized back to FP32 format during the forward pass.
Therefore, the layer computations are performed in FP32 format using exact multipliers(EXf32), as mentioned in Table~\ref{tab:comparison}, which allows us to avoid \gls{lut}-based emulation of approximate multipliers during retraining.
Moreover, the gradients are computed in FP32 format as usual, and the weights are updated to the FP32 format as well during the backward pass instead of going through the quantization process (see the black and blue arrows in Figure~\ref{fig:patqat_framework}), which is different from the methods used in the literature~\cite{yu_toward_2024}.
Our current retraining method supports convolutional (CONV) and fully-connected/linear (FC) layers.
For CONV layers, we apply per-filter INT8 quantization to quantize the weights, while for FC layers we apply per-layer INT8 quantization.
Such selection of quantization methods depends on the inference framework, TFLite, see Section~\ref{sec:model_conversion}.
We use the torch.round~\cite{paszke2017pytorch} function to round the INT8-quantized weights to the nearest pattern INT8 value.

Our defined patterns in Table~\ref{tab:pattern_analysis} include both exact and approximate regions of an AMi8 multiplier; the retraining process allows the weights to be adapted to values that may produce approximation error during inference. 
But the accuracy loss is mitigated since most of the weights are still within the exact region of the AMi8.
Moreover, a pattern rounding technique is adopted during inference to avoid the accuracy loss due to approximation error (see Section~\ref{sec:Model_Deployment}). 
The key reason for including the approximate region (i.e., 127 \new{or both -127 and 127}) in the pattern is to maintain the symmetric requirement of INT8 quantization for the weights, which is also dictated by the inference framework used, TFLite (see Section~\ref{sec:model_conversion}).
In addition, symmetric quantization reduces the quantization parameters and simplifies the hardware design of the accelerator~\cite{jacob_2018_CVPR}.
Since TER~\cite{yu_toward_2024} defines the pattern as the exact region of an AMi8, due to the asymmetric nature of the exact region, the retraining process in TER does not satisfy the symmetric quantization requirement for the weights, which limits the accuracy recovery of TER for our selected AMi8 (see Section~\ref{sec:compare_retraining}).

Any layer with this patterned INT8 quantization function during the forward pass is renamed to patterned-quantized layer in the model architecture graph, e.g., a CONV layer becomes a patternQuant layer.
During retraining, the weights of the first CONV and last FC layers are kept in INT8-quantized format, since previous work~\cite{kim_effects_2022} shows that keeping these layers in higher regular INT8 format helps in improving accuracy.
Retraining is performed for 20 epochs with an initial learning rate of 0.001, which is reduced by a factor of 10 after 5 and 15 epochs.
A \gls{sgd} optimizer with momentum of 0.9 and weight decay of 1e-4 is used for retraining, along with a batch size of 512.
Note that although we perform retraining using PyTorch, the TensorFlow framework can also be used for retraining with the same pattern-guided quantization method, which is one of the advantages of our proposed method compared to prior work~\cite{danopoulos_adapt_2023,pinos_acceleration_2023}.



\section{\texttt{FAME} platform} 
\label{sec: method}



\new{The goal of \texttt{FAME} is to use the flexible logic fabric of FPGAs to implement approximate multipliers on a custom accelerator and evaluate application-level accuracy of INT8-quantized \glspl{dnn}.
For this purpose, the \gls{tflite}~\cite{tflite} framework is used to deploy the DNN models on the designed accelerators.
Although \texttt{FAME} is designed to work standalone without \texttt{PatQAT}, Figure~\ref{fig:proposed_framework} illustrates the integration of \texttt{FAME} with \texttt{PatQAT} to demonstrate the complete evaluation process from retraining to inference evaluation on the target AMi8-based accelerators.
The integrated framework has three main stages: model retraining using \texttt{PatQAT} (blue), model conversion (yellow), and model evaluation with \texttt{FAME} on the target AMi8-based accelerators (green).
\texttt{FAME} has two sub-components: accelerator design and model deployment.
While \texttt{PatQAT} is described in Section~\ref{sec:patqat}, this section describes the other two stages of the integrated framework, i.e., model conversion and deployment on the designed accelerator.}


\subsection{Model Conversion}
\label{sec:model_conversion}

To execute a PyTorch-trained \gls{dnn} model on \gls{tflite}, it must first be converted into a \gls{tflite} format.
A pre-trained FP32 model from Torchvision~\cite{pytorchvision} (i.e., a library of PyTorch) can directly be converted using the Nobuco~\cite{nobuco} converter from a PyTorch to a TensorFlow~\cite{abadi2016} model.
However, for a retrained model using \texttt{PatQAT} in a PyTorch environment, CONV layers in the model graph are replaced with patternQuant layers.
The forward function of the patternQuant layers include pattern-guided INT8 quantization (Figure~\ref{fig:patqat_framework}). Since patternQuant layers are custom and include pattern-guided INT8 quantization, Nobuco~\cite{nobuco} does not support the conversion of such layers from PyTorch to TensorFlow.
To solve this, we first replace patternQuant layers in the retrained model graph with CONV layers using their dequantized float32 weights. 
We call this model graph modification.

After this, the TensorFlow model is obtained using Nobuco~\cite{nobuco} and then converted to \gls{tflite} format using the \gls{tflite} converter~\cite{tflite}.
The \gls{tflite} converter used \gls{ptq} to quantize the model to the INT8 format for deployment on the designed accelerators that include AMi8.
The activation and weights are quantized to INT8 using the default symmetric affine quantization method of the \gls{tflite} converter~\cite{jacob_2018_CVPR}.
Since the dequantized FP32 weights of retrained models represent the defined patterned INT8 values, the affine quantization of the \gls{ptq} process basically maps the weights back to the same patterned INT8 values, as shown in Figure~\ref{fig:patqat_framework}.
Additionally, activations are calibrated to INT8 using 1000 random ImageNet~\cite{krizhevsky2012imagenet} training samples via the default method of \gls{tflite} converter~\cite{tflite}. 
This slightly reduces \gls{tflite} inference accuracy compared to the retrained PyTorch model (Section~\ref{sec:eval_exact_mult}).


\subsection{Accelerator Design}

To design an FPGA-based accelerator, we used the SECDA~\cite{haris_secda_2021} methodology and modified the \gls{vmac} proposed in the SECDA-TFLite~\cite{haris_secda_tflite_2023} toolkit.
SECDA-TFLite takes a \gls{tflite} model as input and runs different tools (e.g., benchmarking) on custom FPGA-based accelerators in a heterogeneous way along with the CPU.
\newcmready{Note that the accelerator is not the primary contribution of this paper; rather, the contribution lies in using it to enable efficient accuracy evaluation of AMi8-based DNNs.}

One of the key modifications in the \gls{vmac} design is increasing the number of \gls{gemm} units from 4 to 8 to increase the throughput of the accelerator as well as utilizing more FPGA resources of the target FPGA device, the KRIA~\cite{KRIA} board.
Note that the \gls{vmac} design in SECDA-TFLite targets a smaller FPGA device, Zynq-7000~\cite{PYNQz2}, hence only 4 \gls{gemm} units were implemented to fit within the resource budget of that device.
In this scaling process, we also utilized the \gls{uram} resources of the KRIA to implement 4 more \gls{gemm} units. 
Another key modification is replacing the EXi8 in the \gls{gemm} units with the AMi8 described in Section~\ref{sec:ami8}.

Each approximate multiplier is implemented in Verilog (\texttt{approx\_mult.v}).
Within SECDA-TFLite, the hardware generator module generates the \gls{hdl} code of the accelerator from SystemC to Verilog \gls{hdl} using Vivado \gls{hls}~\cite{amd_vivado_ipi_2019}.
For the \gls{gemm} unit design of the accelerator, we instantiate INT8 multipliers with inline off \gls{hls} \textit{pragmas} which create separate Verilog modules for INT8 multipliers. 
We modify the hardware generator module file in such a way that it replaces the regular INT8 multipliers (\texttt{mul\_s8.v}) with our approximate multiplier Verilog file (\texttt{approx\_mult.v}) after the \gls{hls} \gls{ip} core generation is done. 
The hardware generator later uses this modified \gls{hls} \gls{ip} core within a full hardware design that includes the Zynq UltraScale+ MPSoC and \gls{dma} \glspl{ip} to generate the final hardware files (i.e., the bitstream). 
\new{This whole hardware generation process is automated using a Python file within SECDA-TFLite and modified accordingly to include the targeted approximate multipliers in the accelerator design.
In this way, we can generate 27 different accelerator designs for the 27 different approximate multipliers mentioned in Section~\ref{sec:ami8}.}
The accelerator generation is one time process, and the generation time varies depending on the host machine. 
\new{For our host machine with 12th Gen Intel® Core™ i7-12700 CPU and 64GB RAM, the generation time is around 25 minutes for each accelerator design.
After generating the hardware files, we can deploy the target DNN models on the generated accelerators and evaluate the accuracy without any additional hardware generation process.}

\begin{figure*}[t]
    \centering
    \includegraphics[width=0.9\textwidth]{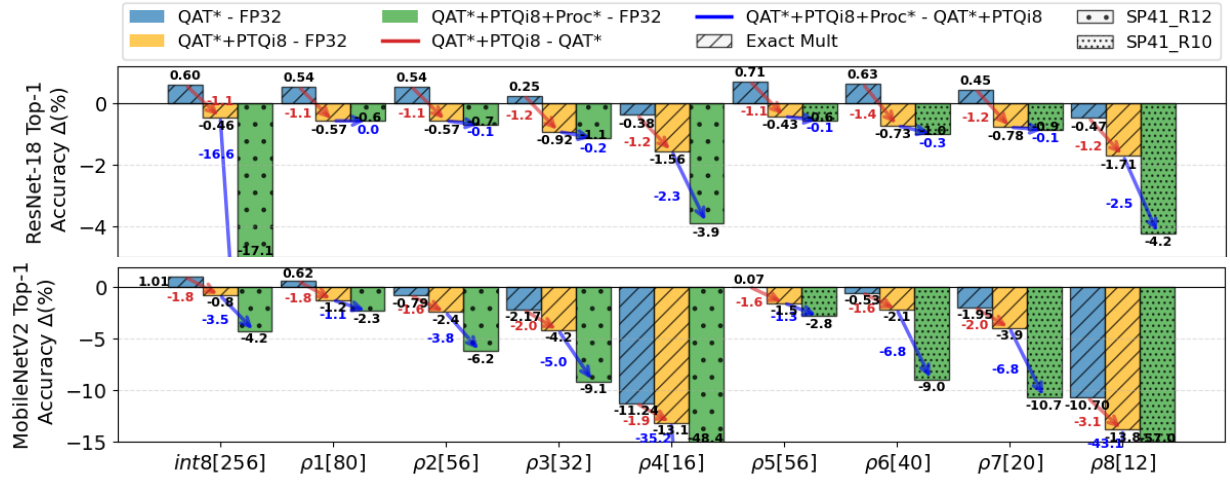}
    \caption{\new{Top-1 Accuracy(\%) Difference Analysis vs retraining; QAT*: QATi8 for INT8 and \texttt{PatQAT} for $\rho1$-$\rho8$; Proc*: without \texttt{PatProc} for INT8 and \texttt{PatProc} for $\rho1$-$\rho8$; FP32 Top-1 Accuracy: ResNet18 (69.76\%) and MobileNetV2 (72.15\%). }}
    \label{fig:top1_diff}
\end{figure*}


\subsection{Model Deployment}
\label{sec:Model_Deployment}


To deploy a converted \gls{tflite} model on a designed accelerator, we need to modify the application code and delegate code that connects the designed accelerator as a custom hardware backend within SECDA-TFLite. 
The delegate system allows to offload certain layers of a TFLite model to custom hardware backends for acceleration.
In our case, only the layers that were quantized to patterned INT8 values are offloaded to the designed accelerator and the rest of the layers, including first CONV and last FC layers, run on the CPU using INT8 weights.
The delegate code of our accelerator maintains such a selection of layers to offload to the accelerator.
Within our delegate code, we also pre-process the patterned INT8 weights of pattern-quantized layers, named as \texttt{PatProc}.
The \texttt{PatProc} rounds the patterned INT8 weights to the defined range of the exact region of the target AMi8 (see Table~\ref{tab:pattern_analysis}).
This rounding introduces a Pattern Rounding Error (\texttt{PRE}) during inference to avoid the approximation error due to multiplication for the target AMi8.
For example, for $\rho$1, all the 127 values in the weights are rounded to 120.
Note that the \texttt{PatProc} step is done only once before the inference starts and does not add any overhead during the inference.


To evaluate the accuracy of a \gls{tflite} model on the ImageNet dataset~\cite{krizhevsky2012imagenet}, we modified the ImageNet Image Classification application code within SECDA-TFLite to use our custom accelerator delegate.
We modified the input image pre-processing step to use the same processing as the one used during retraining in PyTorch.
This modification is necessary to maintain the trained accuracy during inference~\cite{louloudakis_fetafix_2025}.
The ImageNet Image Classification application enables us to run the 50K test images of the ImageNet~\cite{krizhevsky2012imagenet} dataset and evaluate the Top-1 to Top-10 accuracy (\%) of the DNN models on the designed accelerators with approximate multipliers.

\newcmready{
The accuracy of an INT8-QAT TFLite model that is prepared without \texttt{PatQAT} (i.e., weights are quantized to any INT8 values) can be independently evaluated on \texttt{FAME} using an AMi8-based accelerator along with ImageNet-based application, without requiring any \texttt{PatProc} step.
The \texttt{PatProc} step becomes necessary only when evaluating a \texttt{PatQAT} model on \texttt{FAME} as mentioned in previous paragraph.}


\section{Evaluation}

\newcmready{In this section, we first use \texttt{FAME} to evaluate DNN models, i.e., Top-1 Accuracy(\%), retrained with either \texttt{PatQAT} or standalone INT8 QAT (i.e., weights are quantized to any INT8 values) on two representative AMi8-based accelerators (SP41 from the SPRIM8(R12) and SPRIM8(R10)), as shown in Figure~\ref{fig:top1_diff}.
Evaluating standalone INT8 QAT establishes the baseline for measuring the accuracy improvement of \texttt{PatQAT}, while demonstrating that \texttt{FAME} is independent of the retraining method.
To isolate the impact of each stage, we evaluate model accuracy after retraining (Figure~\ref{fig:proposed_framework}, blue) and model conversion (yellow) using exact multipliers, and evaluate the converted TFLite models on the target AMi8-based accelerator using \texttt{FAME} (green).
The same color coding is used in Figure~\ref{fig:top1_diff}, which shows the accuracy changes across these stages for \texttt{PatQAT} and compares them with standalone INT8 QAT for the two selected AMi8. 
Table~\ref{tab:acc_results} reports the converted TFLite model accuracy (bolded) for all 27 AMi8 across all patterns ($\rho$1--$\rho$13) for the target DNNs.
Next, we evaluate hardware metrics such as \gls{pdp} in relation to accuracy to identify the optimal AMi8 design. 
Finally, we compare our retraining and inference evaluation methodology with previous work.}

\subsection{Experimental Setup}

For retraining with \texttt{PatQAT}, we use the ImageNet dataset~\cite{krizhevsky2012imagenet} and two popular CNN models, ResNet-18~\cite{He2016DeepRL} and MobileNetV2~\cite{mbv2} from Torchvision~\cite{pytorchvision}.
We used PyTorch~\cite{paszke2017pytorch} and an NVIDIA RTX A6000 GPU with 48GB of memory for all retraining experiments. 
To evaluate a DNN model after conversion and \gls{ptq}, we used a 12th Gen Intel® Core™ i7-12700 CPU with 64GB RAM using the TFLite~\cite{tflite} framework.
For accuracy evaluation after \texttt{PatProc}, we used the Kria KV260~\cite{KRIA} board, which features a Xilinx Zynq UltraScale+ MPSoC integrating a quad-core ARM Cortex-A53 CPU, programmable logic fabric, and 4GB of DDR3 memory.


\begin{table*}[t]
\Large
\centering
\caption{\new{Evaluation with \texttt{FAME} for SPRIM8 and EVO~\cite{mrazek2017evoapprox8b} Class AMi8 for ImageNet-based DNNs: Top-1 Accuracy (\%).}}
\label{tab:acc_results}
\resizebox{1.0\textwidth}{!}{%
\begin{tabular}{ll|cccccccccc||cccccccccc}
\hline
 \multicolumn{2}{c|}{\textbf{DNN}} & \multicolumn{10}{c||}{\textbf{ResNet-18}} & \multicolumn{10}{c}{\textbf{MobileNetV2}} \\
 \cline{1-22}
 \multicolumn{2}{c|}{\textbf{VMAC}} & SP41 & SP42 & SP43 & SP51 & SP52 & SP53 & SP44 & SP54 & SP45 & SP55 & SP41 & SP42 & SP43 & SP51 & SP52 & SP53 & SP44 & SP54 & SP45 & SP55 \\
\hline
\multirow{4}{*}{\rotatebox[origin=c]{90}{\shortstack{SPRIM8\\(R12)}}}
             & $\rho_1$ & \textbf{69.19} & \textbf{69.19} & \textbf{69.19} & 57.90 & 57.90 & 57.90 & 0.098 & 0.10 & 0.086 & 0.09  & \textbf{69.86} & \textbf{69.86} & \textbf{69.86} & 66.37 & 66.37 & 66.37 & 0.09 & 0.08 & 0.10 & 0.10 \\
             & $\rho_2$ & 69.05 & 69.05 & 69.05 & \textbf{69.04} & \textbf{69.04} & \textbf{69.05} & 0.17 & 0.17 & 0.10 & 0.11    & 65.95 & 65.95 & 65.95 & \textbf{65.95} & \textbf{65.95} & \textbf{65.95} & 0.05 & 0.05 & 0.12 & 0.12 \\
             & $\rho_3$ & 68.63 & 68.63 & 68.63 & 68.63 & 68.63 & 68.63 & \textbf{68.63} & \textbf{68.63} & 0.16 & 0.16           & 63.03 & 63.03 & 63.03 & 63.03 & 63.03 & 63.03 & \textbf{63.03} & \textbf{63.03} & 0.10 & 0.10 \\
             & $\rho_4$ & 65.88 & 65.89 & 65.89 & 65.88 & 65.89 & 65.89 & 65.90 & 65.89 & \textbf{65.89} & \textbf{65.89}         & 23.77 & 23.77 & 23.77 & 23.77 & 23.77 & 23.77 & 23.77 & 23.77 & \textbf{23.77} & \textbf{23.77} \\
\hline
\multirow{4}{*}{\rotatebox[origin=c]{90}{\shortstack{SPRIM8\\(R10)}}}
             & $\rho_5$ & \textbf{69.20} & \textbf{69.20} & \textbf{69.20} & 66.99 & 66.99 & 66.99 & 0.13 & 0.11 & 0.08 & 0.07    & \textbf{69.36} & \textbf{69.36} & \textbf{69.36} & 68.50 & 68.50 & 68.50 & 0.07 & 0.07 & 0.13 & 0.13 \\
             & $\rho_6$ & 68.77 & 68.77 & 68.77 & \textbf{68.77} & \textbf{68.77} & \textbf{68.77} & 3.49 & 3.49 & 0.06 & 0.06    & 63.17 & 63.17 & 63.17 & \textbf{63.17} & \textbf{63.17} & \textbf{63.17} & 0.22 & 0.22 & 0.12 & 0.12 \\
             & $\rho_7$ & 68.90 & 68.90 & 68.90 & 68.90 & 68.90 & 68.90 & \textbf{68.90} & \textbf{68.90} & 0.11 & 0.11           & 61.45 & 61.45 & 61.45 & 61.45 & 61.45 & 61.45 & \textbf{61.45} & \textbf{61.45} & 0.08 & 0.08 \\
             & $\rho_8$ & 65.52 & 65.52 & 65.52 & 65.52 & 65.52 & 65.52 & 65.52 & 65.52 & \textbf{65.52} & \textbf{65.52}         & 15.18 & 15.18 & 15.18 & 15.18 & 15.18 & 15.18 & 15.18 & 15.18 & \textbf{15.18} & \textbf{15.18} \\
\hline
\end{tabular}%
}

\vspace{0.4em}


\resizebox{0.77\textwidth}{!}{%
\begin{tabular}{ll|ccccccc||ccccccc}
\hline
 \multicolumn{2}{c|}{\textbf{VMAC}} & 1KVM & 1KV8 & 1KV9 & 1KVP & 1KVQ & 1KVA & 1KX5 & 1KVM & 1KV8 & 1KV9 & 1KVP & 1KVQ & 1KVA & 1KX5 \\
\hline
\multirow{5}{*}{\rotatebox[origin=c]{90}{\shortstack{EVO\\(MUL8S)}}}
             & $\rho_9$    & \textbf{69.96} & 69.45 & 58.19 & 58.19 & 0.46 & 0.46 & 0.12              & \textbf{71.63} & 71.60 & 69.81 & 69.81 & 51.02 & 51.02 & 0.12 \\
             & $\rho_{10}$ & 69.74  & \textbf{69.74} & 68.44 & 68.44 & 26.75 & 26.75 & 0.14           & 71.44 & \textbf{71.44} & 70.81 & 70.81 & 63.45 & 63.45 & 3.08 \\
             & $\rho_{11}$ & 69.45  & 69.45 & \textbf{69.45} & \textbf{69.45} & 61.71 & 61.71 & 0.19  & 69.44 & 69.44 & \textbf{69.44} & \textbf{69.44} & 66.88 & 66.88 & 29.24 \\
             & $\rho_{12}$ & 69.26  & 69.26 & 69.26 & 69.26 & \textbf{69.26} & \textbf{69.26} & 35.43 & 62.48 & 62.48 & 62.48 & 62.48 & \textbf{62.48} & \textbf{62.48} & 52.74 \\
             & $\rho_{13}$ & 66.20  & 66.20 & 66.20 & 66.20 & 66.20 & 66.20 & \textbf{66.20}          & 16.69 & 16.69 & 16.69 & 16.69 & 16.69 & 16.69 & \textbf{16.69} \\
\hline
\end{tabular}%
}
\end{table*}

\subsection{DNN Accuracy Evaluation}


\subsubsection{Using Exact Multipliers}
\label{sec:eval_exact_mult}

The blue bars in Figure~\ref{fig:top1_diff} show the Top-1 accuracy difference of retrained models (QAT*) with respect to the pre-trained FP32 models.
The baseline INT8 model is retrained in a similar way to the \texttt{PatQAT} models, but with the weights quantized to any INT8 values without any pattern constraint while keeping the activations in FP32 precision. 
The subsequent yellow bars show the Top-1 accuracy difference of the converted TFLite models (i.e., QAT*+PTQi8, see Section~\ref{sec:model_conversion}) with respect to the pre-trained FP32 models. 
The TFLite conversion includes the calibration of activations to INT8 precision using PTQ, which is expected to cause accuracy loss.
The red arrows show the Top-1 accuracy difference between retraining and model conversion stages, which represents the accuracy loss due to the calibration of activations to INT8 precision during TFLite conversion, since other sources of error (e.g., input data preprocessing) during model conversion are minimized.

\new{When we compare the retraining stage (i.e., blue bars) comparing to the baseline INT8 model, we can see that the accuracy gain diminishes gradually as the pattern size decreases for both ResNet-18 and MobileNetV2, which is expected since the smaller pattern size indicates a greater quantization error during retraining.
Within $\rho$1 to $\rho$8 (blue bar), ResNet-18 performs well across all patterns with an average accuracy gain of 0.28\% compared to the pre-trained FP32 model.}
Even when the pattern ($\rho$8) size is limited to only 12 values, the accuracy loss is only 0.47\%.
Another interesting observation is that $\rho$5 achieves better accuracy gain compared to $\rho$2, even though both patterns are of size 56.
This indicates that the specific pattern values also play an important role, in addition to the pattern size.
This conclusion can be further reinforced by comparing $\rho$5[56] with $\rho$1[80] or $\rho$6[40] with $\rho$1[80].
In both cases, the smaller pattern size achieves better accuracy gain. 
In contrast, MobileNetV2, within $\rho$1 to $\rho$8 (blue bar), shows that the accuracy is more sensitive to the pattern size.
Only $\rho$1 with size 80 achieves an accuracy gain of 0.62\%, and $\rho$5 with size 56 achieves an accuracy gain of 0.07\%.
However, the comparison between $\rho$2 and $\rho$5, for size 56, is also interesting for MobileNetV2, as $\rho$5 achieves 0.72\% better accuracy compared to $\rho$2.
Therefore, we can conclude that although the pattern size is more important for accuracy for MobileNetV2, the specific pattern values also play an important role.

\new{According to the red arrows (i.e., difference between blue and yellow bar) in Figure~\ref{fig:top1_diff}, the accuracy loss due to the calibration of activations to INT8 precision during TFLite conversion is on average 1.19\% for ResNet-18 and 1.95\% for MobileNetV2 across all patterns, which is inline with the baseline INT8 model with 1.1\% accuracy loss for ResNet-18 and 1.8\% for MobileNetV2. 
This indicates that the accuracy loss due to the calibration of activations to INT8 precision during TFLite conversion is similar for both the baseline INT8 and the \texttt{PatQAT} models.}

\subsubsection{Using AMi8 with \texttt{FAME}}

The green bars in Figure~\ref{fig:top1_diff} represent the Top-1 accuracy(\%) difference between the TFLite models running on an AMi8-based accelerator during the model deployment stage (QAT* + PTQi8 + Proc*) and the pre-trained FP32 models.
The blue arrows indicate the accuracy difference between model conversion and the model deployment stages. 
Since we isolated the accuracy loss due to the calibration of activations to INT8 precision during model conversion, the accuracy loss in this case is solely due to the respective error of an AMi8-based accelerator.
In the model deployment stage, we use \texttt{PatProc} for the evaluation of retrained models with $\rho_1$ to $\rho_8$ patterns, whereas for the baseline INT8 model, \texttt{PatProc} is not required since weights can take any INT8 value.
Therefore, the accuracy loss in the case of the baseline INT8 model on an AMi8-based accelerator is solely due to the approximation error of an AMi8.

During model deployment for $\rho_1$ to $\rho_8$, \texttt{PatProc} rounds the +127 values in PTQ INT8 weights to the nearest pattern value (see Section~\ref{sec:Model_Deployment}), introducing a pattern rounding error (\texttt{PRE}). 
The magnitude of \texttt{PRE} depends on the highest INT8 value in the exact region of a given pattern (see Table~\ref{tab:pattern_analysis}). 
For instance, \texttt{PRE} is smaller for pattern $\rho_1$ than for $\rho_4$, since +127 is rounded to +120 in $\rho_1$, whereas it is rounded to +97 in $\rho_4$.
Consequently, patterns with larger \texttt{PRE} incur higher accuracy loss, as indicated by the blue arrows in Figure~\ref{fig:top1_diff} for $\rho_1$ to $\rho_8$.
For both ResNet-18 and MobileNetV2, patterns $\rho_4$ and $\rho_8$, which exhibit higher \texttt{PRE}, result in greater accuracy degradation compared to other patterns.



Due to approximation error, INT8 quantized ResNet-18 suffers a 17.1\% accuracy loss relative to pre-trained FP32, reduced to 0.6\% with \texttt{PatQAT} and \texttt{PatProc} for $\rho_1$ and $\rho_5$.
For MobileNetV2, $\rho_1$ achieves a 2.3\% loss, compared to 4.2\% for INT8 quantization.
In both cases, \texttt{PatQAT} constrains weights to the pattern, while \texttt{PatProc} mitigates approximation errors, reducing the accuracy loss relative to the pre-trained FP32 model.

\new{In Table~\ref{tab:acc_results}, the Top-1 accuracy for a given pattern is identical across multiple approximate multipliers. 
For example, for ResNet-18 with Pattern ($\rho$1), the Top-1 accuracy is 69.19\% for SP41--SP43 in the SPRIM8 (R12) multiplier class. 
This can be explained by the fact that approximate multipliers sharing the same exact region produce identical accuracy results.
Table~\ref{tab:pattern_analysis} illustrates how approximate multipliers are grouped according to their exact regions. 
Since \texttt{PatProc} constrains the weights to values within the exact region of a selected pattern, the corresponding approximate multipliers behave identically to exact multipliers for those values and therefore introduce no approximation error. 
To highlight this property, groups of approximate multipliers with the same exact region are indicated in Table~\ref{tab:acc_results} by bolding the corresponding Top-1 accuracy values for both ResNet-18 and MobileNetV2. 
Furthermore, owing to the hierarchical nature of the patterns shown in Table~\ref{tab:pattern_analysis}, the Top-1 accuracy for the smallest pattern (e.g., $\rho$4) is identical across all approximate multipliers within the SPRIM8 (R12) class. 
This hierarchical relationship can also be observed in Table~\ref{tab:acc_results} by examining the accuracy values to the left of the bolded entries for a given pattern, DNN model, and approximate multiplier class.
Note that Table~\ref{tab:acc_results} reports the Top-1 accuracy for all 27 AMi8 designs within a selected pattern, DNN model, and approximate multiplier class to illustrate the hierarchical relationship between patterns and the grouping of multipliers with identical exact regions. 
In a practical deployment scenario, however, it is sufficient to evaluate only the target AMi8 design associated with a given pattern to obtain the Top-1 accuracy for that pattern (i.e., the bolded value in Table~\ref{tab:acc_results}).
}

\begin{table*}[t]
\small
\centering
\caption{\new{Inference evaluation comparison with prior work. $T_{ax}$ (ms/image): time per image using approximate multipliers; Avg Power (W); Perf./W ((image/s)/W); $T_{ex}$ (ms/image): time per image using exact multipliers; Overhead ($\times$): $T_{ax}/T_{ex}$.}}
\label{tab:inference_evaluation}
\resizebox{1.0\textwidth}{!}{
\begin{tabular}{l l l c l r r r r r}
\toprule
\multicolumn{1}{c}{\multirow{2}{*}{\textbf{DNN}}} & \multicolumn{1}{c}{\multirow{2}{*}{\textbf{Work}}} & \multicolumn{1}{c}{\multirow{2}{*}{\textbf{Hardware}}} & \multicolumn{1}{c}{\multirow{2}{*}{\textbf{Batch}}} & \multicolumn{4}{c}{\textbf{AMi8}} & \multicolumn{1}{c}{\multirow{2}{*}{\textbf{$T_{ex}$}}} & \multicolumn{1}{c}{\multirow{2}{*}{\textbf{Overhead}}} \\
\cmidrule(lr){5-8}
 &  &  &  & \textbf{Type} & \textbf{$T_{ax}$} & \textbf{Avg Power} & \textbf{Perf./W} &  &  \\
\midrule
\multicolumn{1}{c}{\multirow{3}{*}{ResNet-18}} & AdaPT~\cite{danopoulos_adapt_2023}        & CPU-i5 3rd gen, RAM 4GB & 64 & LUT-AMi8 & 357.76 & 38.27 & 0.07 & 34.4 & 10.41$\times$ \\
                                               & TFApproxIL~\cite{pinos_acceleration_2023} & Orin Nano, RAM 8GB      & 1  & LUT-AMi8 & 89.0   & 5.98  & 1.88 & 33.0 & 2.70$\times$ \\
                                               & Ours                                      & Kria, RAM 4GB           & 1  & AMi8     & 130.0  & 0.68  & \textbf{11.31} & 130.0 & \textbf{1.00}$\times$ \\
\midrule
\multicolumn{1}{c}{\multirow{3}{*}{MobileNetV2}} & AdaPT~\cite{danopoulos_adapt_2023}        & CPU-i5 3rd gen, RAM 4GB & 64 & LUT-AMi8 & 385.03 & 33.43 & 0.08 & 73.9 & 5.21$\times$ \\
                                                 & TFApproxIL~\cite{pinos_acceleration_2023} & Orin Nano, RAM 8GB      & 1  & LUT-AMi8 & 90.0   & 2.08  & 5.34 & 41.0 & 2.20$\times$ \\
                                                 & Ours                                      & Kria, RAM 4GB           & 1  & AMi8     & 111.0  & 0.48  & \textbf{18.77} & 111.0 & \textbf{1.00}$\times$ \\
\bottomrule
\end{tabular}
}
\end{table*}

\begin{figure}[t]
    \centering
        \includegraphics[width=0.8\columnwidth]{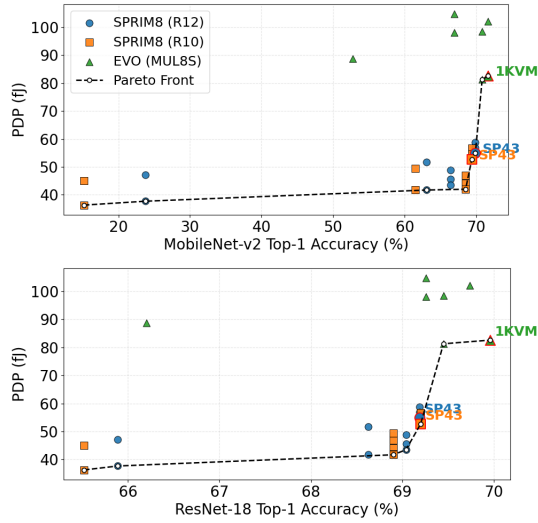}
    \caption{\new{PDP vs. Top-1 Accuracy Comparison. }}
    \label{fig:pdp_acc}
\end{figure}

\subsection{\newcmready{PDP vs. Accuracy}}
\label{sec:hardware_evaluation}

\newcmready{Hardware metrics such as \gls{pdp} indicate the efficiency of an AMi8 design when targeting an ASIC. 
While \texttt{FAME} yields the accuracy of an AMi8-based DNN, \gls{pdp} must  be obtained from ASIC EDA software. 
Combining the two, the \gls{pdp} vs. accuracy comparison identifies the optimal AMi8 design for a given accuracy level.}
\new{The selected 20 AMi8 of SPRIM8 class and 7 AMi8 of EVO class for this paper are originally designed for 45nm and 180nm ASIC implementation, respectively~\cite{Nima-PRIM, mrazek2017evoapprox8b}.
To compare \gls{pdp} of both classes, we estimated the \gls{pdp} of the EVO class multipliers at the 45nm node using the following scaling factor: $\text{Scaling Factor} = \left(\frac{45}{180}\right) \times \left(\frac{1.0}{1.8}\right)^2$, where the feature sizes scale the delay and the power is scaled based on their nominal voltage, 1V and 1.8V for 45nm and 180nm technologies, respectively. This formulation is based on established scaling models where circuit delay scales linearly with the technology feature size, and dynamic power scales quadratically with the nominal supply voltage~\cite{weste2010cmos}.}

\new{Figure~\ref{fig:pdp_acc} illustrates the trade-off between \gls{pdp} and Top-1 accuracy for both ResNet-18 and MobileNetV2 across the evaluated approximate multipliers.
It also highlights the optimal multiplier design for each AMi8 class based on a strict selection policy: we first prioritize maximizing Top-1 accuracy, and then select the design with the lowest \gls{pdp} for a given class and \gls{dnn} model. 
The best Top-1 accuracy obtained across all patterns for each AMi8 design in Table~\ref{tab:acc_results} is used for this comparison.
Following this policy, the SP43 (SPRIM8) and 1KVM (EVO) multipliers were chosen. Compared to other designs in their classes, they delivered superior accuracy owing to their larger pattern sizes ($\rho$1 and $\rho$9, respectively).}

\subsection{Comparison with Prior Work}

\subsubsection{Retraining}
\label{sec:compare_retraining}

\begin{figure}[t]
    \centering
        \includegraphics[width=0.95\columnwidth]{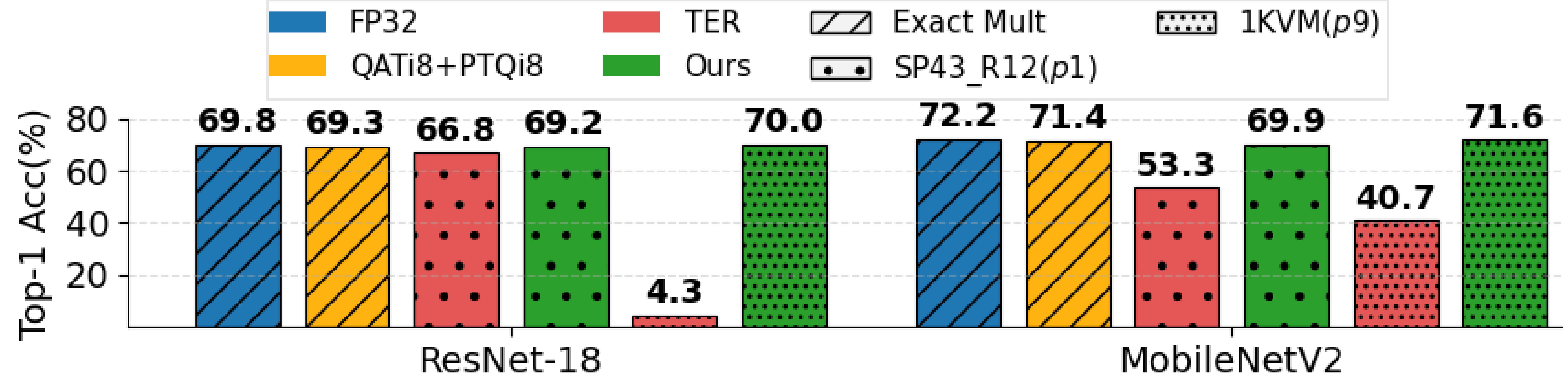}
    \caption{\new{Comparison of \texttt{PatQAT} with TER~\cite{yu_toward_2024} on ImageNet Top-1 accuracy(\%).}}
    \label{fig:accuracy_sota}
\end{figure}

\new{To compare the retraining approach of \texttt{PatQAT} with prior work, we selected TER~\cite{yu_toward_2024} as a representative state-of-the-art method since it also uses exact multipliers during retraining to avoid training-time overhead based on LUT-based approximate multiplier emulation, similar to our approach (see Table~\ref{tab:comparison}).
The best selected AMi8 design in Section~\ref{sec:hardware_evaluation} for each class (i.e., SP43 for SPRIM8 and 1KVM for EVO) is used for this comparison.
We train both \glspl{dnn} for TER~\cite{yu_toward_2024} using the exact region of the AMi8 only and later evaluated using \texttt{FAME}.
As a baseline, we include both FP32 and QATi8+PTQi8 models in the comparison to show the accuracy loss with respect to pre-trained floating-point as well as retrained INT8-quantized models.
Figure~\ref{fig:accuracy_sota} shows the Top-1 accuracy comparison of \texttt{PatQAT} with TER~\cite{yu_toward_2024} for both ResNet-18 and MobileNetV2.
For both \glspl{dnn}, our approach achieves higher Top-1 accuracy (\textbf{on average 28.9\%}) compared to TER~\cite{yu_toward_2024} for the selected approximate multiplier.
Compared to the FP32 baseline, the accuracy loss of the \texttt{PatQAT} approach is on average 0.83\%, whereas the accuracy loss for TER~\cite{yu_toward_2024} is on average 29.7\%, across both \glspl{dnn}.
In addition, a similar trend persists when we compare against the QATi8+PTQi8 baseline.}

The main reason why \texttt{PatQAT} achieves higher accuracy compared to TER~\cite{yu_toward_2024} for the selected approximate multiplier is that \texttt{PatQAT} readjusts the pattern range according to the symmetric INT8 quantization range of $[-127, 127]$ by incorporating a value (e.g., 127) from the approximate region. 
Therefore, patterned INT8 weights map correctly to the dequantized FP32 weights, and later the TFLite converter can quantize the weights to the correct patterned INT8 values.

\subsubsection{Inference Evaluation}

Table~\ref{tab:inference_evaluation} compares the \newcmready{end-to-end} inference efficiency of AdaPT~\cite{danopoulos_adapt_2023}, TFApproxIL~\cite{pinos_acceleration_2023}, and our proposed \texttt{FAME} platform. Unlike AdaPT and TFApproxIL, which rely on \gls{lut}-based software emulation of approximate multipliers on CPU and GPU, respectively, \texttt{FAME} implements AMi8 designs directly in FPGA hardware. 
We evaluate AdaPT~\cite{danopoulos_adapt_2023} on an Intel CPU-i5 3rd Gen with 4GB of memory using the default batch size for both \glspl{dnn}. The 4GB configuration is chosen for a fair comparison with our FPGA-based platform, as the KRIA board also has 4GB of memory, and to reflect similar memory constraints. 
For TFApproxIL~\cite{pinos_acceleration_2023}, we utilize the NVIDIA Jetson Orin Nano (8GB RAM)~\cite{nvidia_jetson_orin_nano_super} as the closest available edge GPU platform. 
\newcmready{Furthermore, with \texttt{FAME}, DNN inference on the KRIA board is performed heterogeneously across the CPU and FPGA: MobileNetV2 uses 53\% of layers on the CPU and 47\% on the FPGA, while ResNet-18 uses 50\% on the CPU and 50\% on the FPGA. 
In both cases, the FPGA executes all CONV layers except the first CONV and depthwise CONV layers, as well as the final FC layer, accounting for $\approx90\%$ of the total computations. 
Profiling shows the CPU-resident layers account for only 29.7\% (MobileNetV2) and 14.6\% (ResNet-18) of end-to-end inference time ($T_{ax}$ in Table~\ref{tab:inference_evaluation}), the rest being spent on the accelerator (including CPU--FPGA data transfers).
The end-to-end inference time thus cannot fall below the execution time of the CPU-resident layers, regardless of the accelerator design.
In contrast, the Orin Nano used by TFApproxIL~\cite{pinos_acceleration_2023} offers 67 TOPS peak~\cite{nvidia_jetson_orin_nano_super} versus the KRIA's 1.4 TOPS~\cite{xilinx_kria_k26_brief_misc} and executes the entire DNN on the GPU.
}


\new{A key advantage of \texttt{FAME} is that the inference time using approximate multipliers ($T_{ax}$) is identical to that obtained with exact multipliers ($T_{ex}$).
Consequently, \texttt{FAME} introduces \textbf{no evaluation overhead} (1.00$\times$), whereas AdaPT incurs an overhead of 5.21$\times$--10.41$\times$ and TFApproxIL incurs an overhead of 2.20$\times$--2.70$\times$.
This improvement stems from a fundamental difference in how the multipliers are executed. While AdaPT and TFApproxIL use CPU cache locality and GPU parallelism to accelerate \gls{lut}-based emulation, our approach eliminates this overhead by implementing approximate multipliers directly on the FPGA fabric.}

\new{Due to the eliminated emulation overhead, \texttt{FAME} accelerates inference by an average of 3.09$\times$ across both \glspl{dnn} compared to AdaPT ($T_{ax}$).
While \texttt{FAME} achieves 0.75$\times$ the speed of TFApproxIL, this difference stems from the Orin Nano GPU's high peak compute throughput relative to the KRIA \gls{fpga}~\cite{nvidia_jetson_orin_nano_super, xilinx_kria_k26_brief_misc}.
However, this massive compute capability comes at the cost of significantly higher average power consumption.
When factoring in efficiency via the Perf./W metric in Table~\ref{tab:inference_evaluation}, \texttt{FAME} achieves an average of 4.77$\times$ higher performance per watt compared to TFApproxIL.
This strongly indicates that given a device operating within a similar power envelope or compute class as the GPU, an \gls{fpga}-based implementation using \texttt{FAME} would yield superior overall performance compared to its GPU counterpart.}

\new{This result confirms that \texttt{FAME} is a highly efficient platform for exploring large numbers of AMi8 for larger \glspl{dnn} and datasets, as it removes the time-consuming overhead of \gls{lut}-based emulation as seen in prior work.}
Note that the inference time in our platform is influenced by the design of the \gls{vmac}.
Further optimization of the \gls{vmac} architecture is expected to yield additional improvements in evaluation performance.

Note that for inference evaluation we do not take into account the TER~\cite{yu_toward_2024} method because it only considers EXi8 during inference, and that is possible when a \gls{dnn} is retrained with the exact region of an AMi8 only.
Also note that inference evaluation of TER~\cite{yu_toward_2024} is dependent on the retraining technique, whereas our proposed \texttt{FAME} platform is orthogonal to the proposed \texttt{PatQAT} retraining method.
Therefore, using \texttt{FAME} we can evaluate an AMi8-based \gls{dnn}, regardless of whether the \gls{dnn} is retrained or not.
\new{Furthermore, we did not compare the training time with prior work since TER~\cite{yu_toward_2024} already demonstrated that exact multiplier-based retraining incurs significantly lower training time compared to LUT emulation-based retraining using AMi8, and our proposed \texttt{PatQAT} also uses exact multipliers during retraining.}

\section{Conclusion}

We proposed \texttt{FAME}, a novel FPGA-based platform for evaluating \glspl{dnn} employing approximate multipliers, along with a pattern-guided retraining approach, \texttt{PatQAT}, to effectively recover accuracy. \texttt{FAME} enables efficient evaluation across a wide range of approximate multipliers, achieving speedups \new{up to a $3.47\times$} over prior \gls{lut}-based methods by eliminating emulation overhead. Moreover, \texttt{PatQAT} improves model accuracy, achieving an average recovery of \new{28.9\%} compared to a state-of-the-art retraining approach for the evaluated approximate multipliers across both \glspl{dnn}. 
As future work, we plan to extend \texttt{FAME} to support additional types and bit-widths of approximate multipliers, as well as other approximate arithmetic units, such as approximate adders.

\section*{Acknowledgement}
\footnotesize
This work was partially supported by the UK Engineering and Physical Sciences Research Council (grants EP/T517896/1 and EP/W524359/1), the EU project dAIEDGE (GA Nr 101120726), the Innovate UK Horizon Europe Guarantee (GA Nr 10090788), and, for the authors at Heidelberg University, the Hector Stiftung (Project 2304191).



\bibliographystyle{IEEEtran}
\bibliography{references}

\end{document}